\documentclass[aps,pra,tightenlines,12pt,superscriptaddress,notitlepage,nofootinbib,longbibliography]{revtex4-1}

\usepackage{graphicx}  
\usepackage{xcolor}
\usepackage{dcolumn}   
\usepackage{bm}        
\usepackage{amssymb}   
\usepackage{amsmath}
\usepackage{verbatim}
\usepackage[hidelinks]{hyperref}
\usepackage[utf8]{inputenc}     
\usepackage{tabularx}
\usepackage{multirow}
\usepackage[capitalise,noabbrev]{cleveref}
\usepackage[centering,margin=1in]{geometry}
\usepackage{braket}
\usepackage{setspace}

\AtBeginDocument{%
    \newwrite\bibnotes
    \def\bibnotesext{Notes.bib}
    \immediate\openout\bibnotes=\jobname\bibnotesext
    \immediate\write\bibnotes{@CONTROL{REVTEX41Control}}
    \immediate\write\bibnotes{@CONTROL{%
    apsrev41Control,author="08",editor="1",pages="1",title="0",year="1"}}
     \if@filesw
     \immediate\write\@auxout{\string\citation{apsrev41Control}}%
    \fi
}%

\newcommand\snowmass{\begin{center}\rule[-0.2in]{\hsize}{0.01in}\\\rule{\hsize}{0.01in}\\
\vskip 0.1in Submitted to the  Proceedings of the US Community Study\\ 
on the Future of Particle Physics (Snowmass 2021)\\ 
\rule{\hsize}{0.01in}\\\rule[+0.2in]{\hsize}{0.01in} \end{center}}

\begin{document}

\title{Snowmass 2021 White Paper: \\ Tabletop experiments for infrared quantum gravity}

\author{Daniel Carney}
\affiliation{Physics Division, Lawrence Berkeley National Lab, Berkeley, CA, USA}
\author{Yanbei Chen}
\affiliation{Theoretical Astrophysics 350-17, California Institute of Technology, Pasadena, CA, USA}
\author{Andrew Geraci}
\affiliation{Center for Fundamental Physics, Department of Physics and Astronomy,
Northwestern University, Evanston, IL, USA}
\author{Holger M\"uller}
\affiliation{Department of Physics, University of California, Berkeley, CA, USA}
\author{Cristian D. Panda}
\affiliation{Department of Physics, University of California, Berkeley, CA, USA}
\author{Philip C. E. Stamp}
\affiliation{Department of Physics and Astronomy, University of British Columbia, Vancouver, BC}
\affiliation{Theoretical Astrophysics, Cahill, California Institute of Technology, Pasadena CA, USA}
\author{Jacob M. Taylor}
\affiliation{Joint Center for Quantum Information and Computer Science/Joint Quantum Institute,
University of Maryland/NIST, College Park, MD, USA}

\setstretch{1.2}

\begin{abstract}
Progress in the quantum readout and control of mechanical devices from single atoms to large masses may enable a first generation of experiments probing the gravitational interaction in the quantum regime, conceivably within the next decade. In this Snowmass whitepaper, we briefly outline the possibilities and challenges facing the realization of these experiments. In particular, we emphasize the need for detailed theories of modifications to the usual effective QFT of gravitons in the infrared regime $E/L^3 \ll m_{\rm Pl}/\ell_{\rm Pl}^3$ in which these experiments operate, and relations to possible UV completions.
\end{abstract}

\snowmass

\maketitle

\newpage

\tableofcontents

\section{Introduction}

In recent years, building on foundational conceptual proposals \cite{feynman1971lectures,eppley1977necessity,page1981indirect}, a number of authors have discussed the idea of performing lab-scale experiments to probe key questions about the quantum nature of gravity \cite{carney2019tabletop}. These experiments aim to test whether perturbatively quantized general relativity, viewed as an effective quantum field theory \cite{Weinberg:1978kz,Donoghue:1994dn,Donoghue:1995cz,Burgess:2003jk}, correctly describes nature at low energies. A number of alternative scenarios have been proposed. These include models involving the gravitational breakdown of quantum mechanics \cite{diosi1986universal,Penrose:1996cv}, mixed classical-quantum models \cite{ROSENFELD1963353,moller1962theories,Kibble:1979jn}, models of gravity as an emergent force\footnote{Of course, some emergent gravity models reproduce the graviton EFT at long wavelengths \cite{Maldacena:1997re}. However, it is interesting to ask if consistent alternatives could exist.} \cite{Jacobson:1995ab,Verlinde:2010hp}, and ideas about holographic effects in the infrared \cite{chou2017holometer,verlinde2019observational}. Experiments which are conceivably realizable within the next decade can make decisive statements about these models.

In this Snowmass white paper, we briefly outline this emerging research program. We provide a non-exhaustive outlook on theoretical issues and experimental realizations, focusing on major open questions and research opportunities for the next decade. We emphasize, in particular, the need for detailed, theoretically consistent models of infrared gravity which deviate from the standard graviton effective field theory.

\section{Gravitational entanglement and existence of gravitons}

\begin{figure}[t]
\includegraphics[scale=0.9]{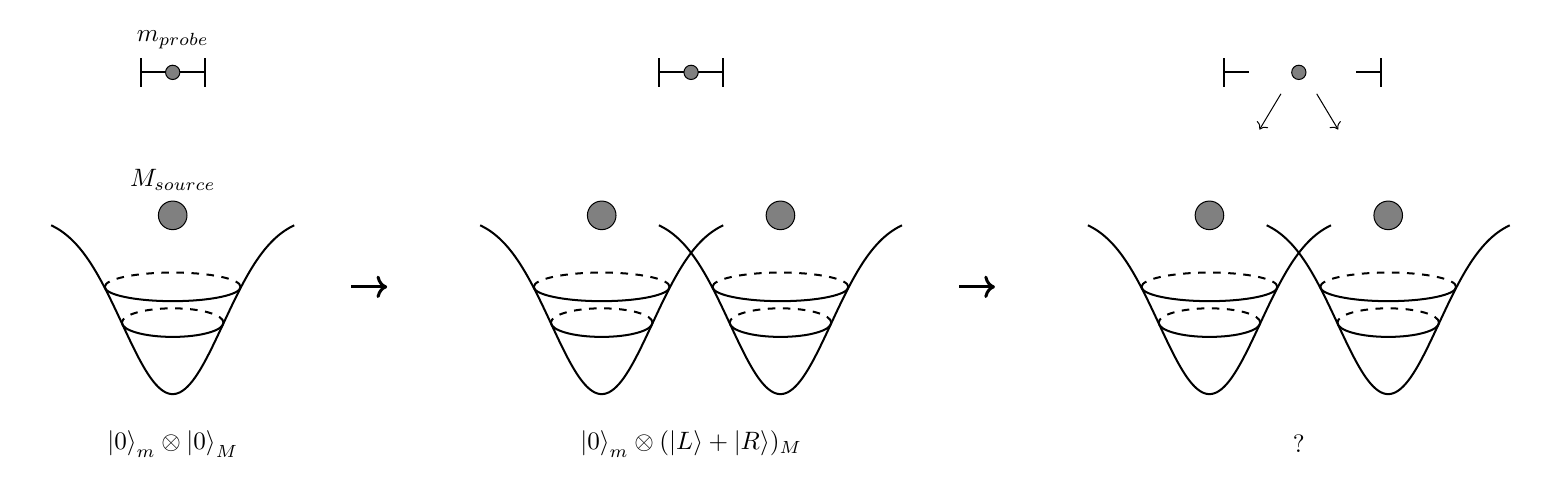}
\caption{Schematic of a gravitational entanglement test. Two masses $M_{\rm source}$ and $m_{\rm probe}$ are prepared in some initial product state. The source mass is then superposed into two locations; if gravity is quantized according to effective field theory rules, the Newton potential part of the gravitational field will likewise become superposed. The probe mass is then allowed to interact and measure this field state, and a Bell-type inequality can be measured to determine if the joint state of the two masses is entangled.}
\label{figure-cartoon}
\end{figure}

An elementary question about quantum gravity is whether the metric is itself a quantized degree of freedom. Dyson and others have argued that it is likely to be impossible to construct a detector sensitive to single graviton events \cite{dysonorig,rothman2006can,dyson2013graviton}. However, there is an alternative path to studying this question, inspired by logic familiar to information theorists: we could test if gravity can produce a state which is provably non-classical, following the classic ideas of Bell \cite{bell1964einstein,aspect1981experimental}. See Figure \ref{figure-cartoon}. Here we discuss both the experimental approaches and interpretation of such tests.

A large number of experimental proposals to test gravity's ability to entangle objects now exist \cite{kafri2013noise,Bahrami:2015wma,anastopoulos2015probing,bose2017spin,marletto2017gravitationally,haine2021searching,Qvarfort:2018uag,Carlesso:2019cuh,PhysRevA.101.063804,howl2021non,Matsumura:2020law,Carney:2021yfw,Carney:2021zax,Streltsov:2021ahn,Pedernales:2021dja,liu2021gravitational,Datta:2021ywm,Feng:2022hfv}. Generally speaking, the major difficulty comes from the need to keep quantum coherence in a massive object over a timescale long enough to see the entanglement. Present proposals generally require using masses at least at the nanogram scale and maintaining coherence times on the order of milliseconds. Some proposals additionally require the ability to initialize at least one of the masses in quantum superposition of two spatially distinct states, presenting a substantial additional challenge with the large masses required. However, while challenging, it is conceivable that one or more of these could be implemented successfully within the next ten years.  

Broadly speaking, there are two main classes of experimental proposals. In both cases, the goal is to rule out any model of gravity which is incapable of generating entanglement between massive objects. The first class of experiments aims to allow a pair of systems to interact gravitational, and use a literal Bell test (or continuous-variable analogue) to demonstrate that the resulting two-body state is entangled. The second class instead aims to rule out some specific behavior in models in which gravity is non-entangling; for example, these models often (and perhaps always) involve anomalous heating and/or decoherence from the gravitational dynamics. To highlight the general ideas, we describe two examples of promising candidate systems for such tests. 

The first involves preparing a massive quantum superposition of a levitated particle to generate a gravitational interaction between itself and a second particle also prepared in a quantum superposition. The phase evolution induced by the gravitational interaction of the two levitated neutral test masses could detectably entangle them via gravity even when they are placed far enough apart to keep other interactions at bay \cite{bose2017spin,marletto2017gravitationally}. Having an embedded spin provides a handle for preparing and manipulating the superposition state along with a possible entanglement witness. In particular, these spin degrees of freedom can be used to perform a CHSH-type Bell test on the masses \cite{bose2017spin}. Such an experiment would require a cryogenic, low-noise environment to suppress additional decoherence mechanisms.

The second approach trades mass for coherence and distance, using new techniques in which atom interferometers have long free evolution times enabled either by microgravity \cite{peters1999measurement} or by trapping atoms during their free fall with potentially macroscopic (mm- to cm-scale) separations \cite{xu2019probing}. In turn, these atoms interact with a massive resonator similar to those proposed in the first approach. The entanglement between the mass and atoms can in principle be detected, and on a more immediate basis, the atoms provide an extremely sensitive probe of any anomalous gravitational heating of the resonator \cite{Carney:2021yfw,Carney:2021zax,Streltsov:2021ahn}. The interaction between the oscillator and the atom in a spatial superposition will weakly modulate the visibility of the atom interferometer, which could be detected in an experiment with high sensitivity and long interaction times. Interrogation times of 20 seconds have been achieved by suspending the spatially separated atomic wave packets in an optical lattice formed by the clean wavefronts of the optical cavity \cite{xu2019probing}.

Many theoretical questions need to be answered in order to fully understand the implications of these experiments. For example, a basic question arises: these experiments probe gravity in the Newtonian limit. If we observe entanglement generation through the Newton potential, is there a precise sense in which this actually demonstrates that the gravitational field is quantized? A number of authors have argued in the affirmative \cite{Belenchia:2018szb,Galley:2020qsf,christodoulou2019possibility,marshman2020locality,Carney:2021vvt,Danielson:2021egj}; see \cite{Anastopoulos:2018drh,marletto2020witnessing,Rydving:2021qua} for some dissenting opinions. In \cite{Carney:2021vvt}, it was demonstrated that observation of non-relativistic gravitational entanglement would require either that quantized gravitational radiation exists or that gravity violates either Lorentz invariance or unitarity. More work on these issues will be of great value, especially in the context of models in which gravity could generate violations of unitarity.

Perhaps more importantly, there is a deep need for detailed, consistent models of gravity coupled to quantum matter in the infrared. Of course, the Einstein equations can be expanded perturbatively and the fluctuations quantized as gravitons, leading to a perfectly well-defined EFT at the energies probed in these experiments \cite{Weinberg:1978kz,Donoghue:1994dn,Donoghue:1995cz,Burgess:2003jk}. If we want to test whether this EFT is the right model of nature, it behooves us to produce alternative models which would produce some other behavior in such experiments. A variety of models of ``classical'' gravity coupled to quantum matter have been produced in recent years, for example \cite{Kafri:2014zsa,tilloy2016sourcing,Hall:2017nzl,Oppenheim:2018igd,Oppenheim:2020ogy,Donadi:2022szl}. These models are constructed precisely to avoid classic no-go theorems of Weinberg and Polchinski \cite{Weinberg:1989cm,Polchinski:1990py} by working entirely in the context of linear quantum mechanics; in the next section, we separately discuss ideas about gravitational breakdown of linear time evolution in the quantum state. Most of these are non-relativistic models; construction of or no-go theorems on non-entangling gravitational models which are fully generally covariant will be extremely valuable. Classification of possible visible signatures would also be very useful; as a concrete example, the classical-quantum model of \cite{Kafri:2014zsa} leads to anomalous heating with a rate proportional to $G_N$, and there may be ways of ruling out such effects which do not require the complete gravitational Bell test-type experiments \cite{Altamirano:2016fas,Carney:2021yfw,Carney:2021zax,Streltsov:2021ahn}.

\section{Gravitational breakdowns of linear quantum mechanics}

Feynman argued as early as the 50's  that one could envisage a breakdown of quantum mechanics, caused by gravity, and that if this occurred, then dimensional arguments would predict it to occur for rest masses of order the Planck mass $m_{\rm pl} \approx 20~{\rm \mu g}$ \cite{feynman57,feynman1971lectures}. He suggested that this breakdown would show up in situations involving delocalized quantum states, in particular for the double-slit system.

Since 1957, several related ideas have emerged. Kibble argued \cite{kibble1} that a low-energy theory of quantum gravity would involve a non-linear brand of QFT, driven by the non-linear metric field dynamics. Many authors have discussed semiclassical gravity, in which the spacetime metric is determined by the expectation value of the stress-energy tensor \cite{kibble2,moller62,rosenfeld63,kiefer07,verdaguer20}.

More recently Penrose \cite{Penrose:1996cv,marshall2003towards} argued that if a quantum superposition involved a massive body in different positions, the ``mismatch'' between the different relevant spacetime metrics would cause a relative phase uncertainty between them, a kind of ``intrinsic decoherence'' in nature. He and others tried to use the non-linear Schrodinger-Newton equation to predict departures from QM caused by Newtonian gravity \cite{moroz1998spherically,tod1999analytical}. In a different development, Diosi and others \cite{diosi,bassi} examined a variant of the ``stochastic collapse" theories pioneered by Ghirardi et al. \cite{ghirardi}; in these ``Continuous Stochastic Localization" (CSL) models, an extraneous noise source associated with Newtonian gravity is supposed to collapse quantum wave-functions. Such an external one-way noise field, a random energy source which acts on the visible universe but is not itself perturbed, makes such theories strictly phenomenological.  Apart from the work of Kibble \cite{kibble1}, none of these analyses are relativistic, nor have they been generalized to relativistic fields.

There has also been criticism of non-linear theories. Some internal inconsistencies of semiclassical gravity have been highlighted (notably by Kibble \cite{kibble1}). Furthermore, when quantum measurements are introduced to non-linear theories of quantum mechanics, the non-collapse of quantum states will violate classical gravity phenomenology~\cite{page1981indirect}, while the collapse of quantum states may lead to superluminal communication \cite{Polchinski:1990py,gisin79,gisin1990nonlocality}. It was shown that in nonlinear theories, different ways of introducing quantum collapse and Born's rule can lead to different phenomenology~\cite{helou2017measurable}, and that some formulations might be immune to superluminal communications~\cite{Kafri:2014zsa,helou2019testing,Kaplan:2021qpv}. 

An example of a consistent nonlinear theory of quantum gravity is
provided by the Correlated Worldline (CWL) theory of Stamp and
collaborators \cite{Barvinsky:2018tsw,Barvinsky:2020pvw,CWL3}. This is a quantum field theory of low-energy gravity in which gravitational correlations between Feynman paths are built in at a fundamental level. These correlations violate the superposition principle and make the theory non-linear; but the theory has been shown to be consistent in that it has a consistent classical limit (Einstein's theory), consistent expansions in $\hbar$ and in $G_N$, and satisfies all the relevant Ward identities \cite{Barvinsky:2018tsw,Barvinsky:2020pvw}. In fact exact solutions can be given for the propagators and equations
of motion in the CWL theory, which allow deviations from quantum mechanics to be calculated with no adjustable parameters \cite{CWL3}; no superluminal communication appears in these solutions. Predictions from this theory for real experiments, when quantum measurement processes are involved, will soon appear.

One of the great attractions of this field is that the theories are within reach of experimental test. At the most massive end, quantum optomechanical techniques allow, in principle, the transfer of the quantum coherence of light to massive mechanical degrees of freedom, up to the 40 kg used in the LIGO gravitational wave detectors \cite{yu-Nat20}. This could potentially be used to demonstrate quantum trajectories and macroscopic quantum entanglement at this scale. At mesoscopic scales, two groups have recently prepared nanoparticles trapped in optical potentials near quantum pure states, observing quantum trajectories of the particles and preparing non-classical out-going light fields~\cite{delic20,magrini2021real,militaru2022ponderomotive}. Rapid progress is now expected in the preparation of much larger masses in non-classical states, as well as in the probing of gravitational coupling between both small and intermediate-sized masses \cite{schmole2016micromechanical,Catano-Lopez:2019yqb,westphal2021measurement}. We expect an explosion of experimental activity in this area in the next few years.

\section{Outlook}

In this brief Snowmass 2021 contribution, we have provided some specific examples of models and tests of the quantum behavior of gravitational interactions that can be probed with realistic, tabletop experiments. There is a clear need for further theoretical work. While a number of models have now been proposed as alternative hypotheses, to be tested against the usual graviton EFT picture, many have crucial open questions about their generally covariant completions, renormalization, and detailed predictions in future experiments. Studies of these models in astrophysical and cosmological scenarios are almost non-existent and would be very interesting. Moreover, while a number of experimental proposals now exist, there is considerable room for new proposals which can alleviate some of the substantial technical challenges involved. This includes both proposals for variants on the above experiment and proposals for testing entirely new predictions from these models, including the graviton effective field theory. We expect continued rapid progress in this emerging research direction over the next decade, and look forward to the first generation of real experimental tests of the quantum nature of gravity.  

\newpage 

\bibliography{snowmass-gravity}

\end{document}